# Prediction of a native ferroelectric metal

**Authors:** Alessio Filippetti,[1,*] Vincenzo Fiorentini,[2,1,*] Francesco Ricci,[2] Pietro Delugas,[3] and Jorge Íñiguez[4,5,*]

**Affiliations:**

[1] CNR-IOM SLACS Cagliari, Istituto Officina dei Materiali, Cittadella Universitaria, Monserrato (CA) 09042-I, Italy.

[2] Dipertimento di Fisica, Università di Cagliari, Cittadella Universitaria, Monserrato (CA) 09042-I, Italy.

[3] Istituto Italiano di Tecnologia IIT, Via Morego 30, 16163 Genova, Italy.

[4] Materials Research and Technology Department, Luxembourg Institute of Science and Technology (LIST), 5 avenue des Hauts-Fourneaux, L-4362 Esch/Alzette, Luxembourg.

[5] Institut de Ciència de Materials de Barcelona (ICMAB-CSIC), Campus UAB, 08193 Bellaterra, Spain.

*Correspondence to: alessio.filippetti@dsf.unica.it, vincenzo.fiorentini@dsf.unica.it,

jorge.iniguez@list.lu

The possibility that metals may support ferroelectricity is an intriguing open issue. Anderson and Blount[1] showed that certain martensitic transitions involve inversion symmetry breaking and, formally, the existence of a polar axis. "Metallic ferroelectric" behavior has thus been claimed for metals undergoing a centrosymmetric (CS) to non-CS structural transformation, such as[2,3] $Cd_2ReO_7$ and $LiOsO_3$, or being natively non-CS, such as[4] $(Sr,Ca)Ru_2O_6$. The same label has been attached to ferroelectric insulators whose polar distortion survives moderate metallicity induced by doping or proximity[5-7]. However, it seems fair to say that none of these systems, nor any other to our knowledge, embodies a truly ferroelectric metal with native switchable polarization and native metallicity coexisting in a single phase. Here we report the first-ever theoretical prediction of such a material. By first-principles calculations, we show that the layered perovskite $Bi_5Ti_5O_{17}$ has a non-zero density of states at the Fermi level and metal-like conductivity, as well as a spontaneous polarization in zero field. Further, we predict that the polarization of $Bi_5Ti_5O_{17}$ is switchable both in principle (the material complies with the sufficient symmetry requirements) and in practice (in spite of being a metal, $Bi_5Ti_5O_{17}$ can sustain a sizable potential drop along the polar direction, as needed to revert its polarization by application of an electric bias). Beyond their conceptual importance, our results reveal striking behaviors – such as the *self screening* mechanism at work in thin $Bi_5Ti_5O_{17}$ layers –

**emerging from the intimate interplay between polar distortions and free carriers. We thus anticipate that our work will have a profound impact in nano-science and nano-technology, as it redefines the frontier of what is possible as regards the control of charges and fields at the nano-scale, with exciting potential applications in areas ranging from photovoltaics to electronics.**

We focus on layered perovskite titanates $A_nTi_nO_{3n+2}$[8,9], whose structure foreshadowing low-dimensional behavior combines with their tunable conduction charge: assuming fixed ionic charges for $A^{3+}$ and $O^{2-}$, Ti has nominal oxidation state of $(3+4/n)$, i.e., between 4+ for n=4 (e.g., the band insulator, high-temperature ferroelectric $La_2Ti_2O_7$) and 3+ in the n→∞ limit (e.g., the Mott-insulating Ti-$d^1$ perovskite $LaTiO_3$). The metallic n>4 phases are not nearly as studied as the end compounds[10-12]. Motivated by experimental reports of possible non-CS structures in the n=5 compound $La_5Ti_5O_{17}$ (La-5517 hereafter)[13], here we discuss this material as well as the alternative composition $Bi_5Ti_5O_{17}$ (Bi-5517).

The n=5 layered titanate can be viewed as a stack of slabs containing 5 [011]-oriented perovskite-like planes and AO terminated (see Fig. 1). (Directions are in the pseudo-cubic setting of the perovskite structure.) The crystal axes are **b**=[011], which will be shown to coincide with the polar axis, and **a**=[100] and **c**=[0-11], which define the plane where the conduction charge is largely confined (see Fig. 2). Our simulation supercell comprises two 5-layer blocks along **b** and is compatible with all structures of interest here. To identify the ground state, we start from the high-symmetry (*Immm* space group, Fig. 1a) structure and condense all its unstable distortions as obtained from a Hessian analysis. For La-5517, the ground state is CS *Pmnn* (Fig. 1b), barring the existence of polarization. We also simulate the experimentally proposed structure of Ref. [13], but find it to be a high-energy unstable configuration.

Inspired by the observation that perovskites where $Bi^{3+}$ replaces $La^{3+}$ tend to be ferroelectric due to $Bi^{3+}$'s tendency to form low-coordination complexes with neighboring oxygens[14,15], we explore symmetry breaking in Bi-5517, obtained by replacing all La atoms with Bi's. As in La-5517, the *Immm* phase is a high-energy saddle point. However, at variance with La-5517, the *Pmnn* structure is also a saddle point for Bi-5517. We then condense the unstable distortions of this phase, and identify as lowest-energy solution a structure with the non-CS *Pm2₁n* space group (Fig. 1c). The symmetry breaking distortion in the *Pm2₁n* phase can be appreciated by looking at the Bi's in the central layer ($Bi_c$ in Fig. 1): while they remain at high-symmetry CS positions in both *Immm* and *Pmnn*, they move off-center in *Pm2₁n*, thus breaking the (011) mirror plane and yielding a symmetrywise ferroelectric structure.

As regards the electronic structure, Bi-5517's *Pm2₁n* phase is clearly metallic, as can be seen in Fig. 2a from the atom- and orbital-resolved density of states (DOS). We have 2 conduction electrons per primitive cell ($3\times10^{21}$ cm$^{-3}$), with E$_F$ crossing the Ti-3d-t$_{2g}$ band manifold approximately 0.4 eV above the conduction band bottom (CBB). The near-CBB DOS highlights a marked two-dimensional character, with 40% of the conduction charge confined within the central Ti layer of each block, 25% in each of the two intermediate layers, and only 5% in each of the edge Ti's. The t$_{2g}$ CBB is split into d$_{yz}$ (laying orthogonally to the stacking plane, rising in energy due to reduced hopping along the stacking direction) and d$_{xy}$/d$_{xz}$ states (the hopping along x being unaffected by the stacking). d$_{xy}$ and d$_{xz}$ are also split: only d$_{xy}$ has significant DOS below E$_F$ in one of the 5-layer blocks, and only d$_{xz}$ in the other, signaling orbital ordering. Fig. [2](b) highlights the anisotropy of the conduction bands: the two occupied bands per block are doubled, there being two blocks in the supercell; yet, the splitting due to inter-block coupling is negligible, confirming good confinement of the conduction electrons within each block. The inset of Fig. [2](b) shows that the bands are completely flat along the Γ-Y (stacking) direction, with no band crossing E$_F$; hence, the system is gapped at Γ along this direction.

Fig. [2](c) shows the Fermi surface (FS). The lowest-energy band S$_1$ consists of two disconnected parallel sheets, and the higher S$_2$ band contributes an elliptic tube. Along Γ-Y (**b** direction) the FS is very flat and resistivity is high, as shown in Fig. 2d. Along Γ-X (**a** direction) the light-mass S$_2$ contributes to mobility, while S$_1$ is disconnected. Finally, along Γ-Z (**c** direction) both sections contribute, but yield relatively low conductivity as they have much heavier mass than along Γ-X. As a result, the conductivity is largely, though not strictly, one-dimensional. The ordering of the conductivities is quite consistent with experiments[9] for La-5517, and the weak insulating upturn observed in La-5517 can be reproduced by inserting small "defect-like" activation energies in the conductivity model (see Methods).

We now tackle the calculation of the ferroelectric polarization appearing in the *Pm2$_1$n* phase of Bi-5517. Let us first note that the polarization – defined as the integrated current flowing along the stacking direction when we move from the CS phase (*Immm*) to the non-CS one (*Pm2$_1$n*) – can be split into contributions from ionic cores, valence electrons and conduction electrons. The first two dominate the effect and are trivial to compute by standard methods[16,17]: we obtain P$_{ion}$=55.5 μC/cm$^2$ and P$_{val}$=-14.6 μC/cm$^2$. In contrast, calculating P$_{cond}$ is not standard. (To our knowledge, the calculation of a polarization in a metal is unprecedented.) Nevertheless, we can take advantage of the two-dimensional localization of the conduction electrons within the slabs of Bi-5517 and implement two independent approaches to calculate P$_{cond}$, obtaining consistent results.

First, we compute P$_{cond}$ from the dipole associated to conduction electrons within a 5-layer block in Bi-5517. Figure [3](#) shows the planar-averaged conduction charge of the *Pm2$_1$n* phase, as well as that of CS system with *Pm2$_1$n* cell parameters and *Immm* atomic positions. The *Pm2$_1$n* phase displays an evident inversion symmetry breaking; a dipole appears within each block and we obtain $P_{cond}$ =-4.0 μC/cm$^2$. Note that a strong 2D charge confinement is apparent in Fig. 3, and this strategy to compute $P_{cond}$ would be exact if the conduction charge were strictly confined within the blocks.

Alternatively, we can compute $P_{cond}$ using a modified version of the Berry phase formalism. Since the occupied conduction bands are flat along the Γ-Y direction of the Brillouin zone (the reciprocal-space signature of confinement) we can generalize the usual formulation to allow for changing numbers of contributing bands on different *k*-point strings (see Methods for details). We eventually obtain $P_{cond}$=-7.5 μC/cm$^2$, which we deem in reasonable agreement with our previous estimate.

Figure 3 also reveals a fascinating effect, namely, how the mobile carriers rearrange *within each of the 5-layer blocks* to screen the field created by the local dipoles resulting from the CS-breaking displacements of the central Bi cations (see the respective enhancement and decrease in electron density on the right and left sides of the Bi$_c$ planes). Remarkably, the ferroelectric instability persists in spite of this *self-screening* mechanism, contradicting the extended notion that an abundance of mobile carriers should prevent any such polar distortion. This result highlights the difference between our material (whose ferroelectric phase has a local, chemical origin associated to the Bi-O bonding) and compounds such as BaTiO$_3$ (where the ferroelectric instability relies on the action of dipole-dipole interactions that are strongly weakened by screening charges).

Hence, our calculations indicate that the *Pm2$_1$n* phase of Bi-5517 has a spontaneous polarization of about 35 μC/cm$^2$, which is in the same league as the most common ferroelectric perovskites (e.g., 30 μC/cm$^2$ for BaTiO$_3$). Ferroelectricity largely originates from Bi$^{3+}$ cations moving off-center in the perovskite framework. This displacement is invertible with respect to the (011) plane, so that switching between two equivalent polar states is in principle possible. The computed ferroelectric well depth of 0.31 meV/Å$^3$ suggests a critical temperature upward of 500 K.

So far we have shown that metallicity coexist with zero-field polarization in Bi-5517. Is it possible to switch the polarization of this ferroelectric metal? Or, put differently, can a finite Bi-5517 sample sustain a finite field, as would be required to switch its polarization? For metallic contacts to Bi-5517, one may expect the bias to induce a current rather than to act on the polar distortion. For Bi-5517 cladded within insulating layers, current flow is precluded by construction (neglecting tunneling), but mobile carriers should screen an applied bias, and leave the CS-breaking distortion unaffected. Nevertheless, as it turns out, Bi-5517 is quite at odds with this reasonable expectation. We show this studying a superlattice (SL) of alternating Bi-5517 (one primitive cell, ~31 Å thick) and Bi$_2$Zr$_2$O$_7$ (BZO-227, n=4 of the same family; layer ~26 Å thick) layers. BZO-227 acts as a cladding insulator providing seamless stoichiometric continuity on the A-cation site as well as effective confinement of the conduction electrons within Bi-5517. We compare a SL where Bi-5517 is non-CS (starting from the *Pm2₁n* bulk phase) with a reference SL which is a suitable symmetrization of the non-CS one (which yields a *Pmnn*-like structure for the Bi-5517 layer).

Figure 4 shows planar and filter averages[7] (see Methods) of the potentials and conduction densities of the two SLs, and their differences. The key result is the sizable depolarizing field $E_{dep}$~20 MV/m=0.02 GV/m (which we estimate from the potential slope in the central region of Bi-5517; see Fig.4, bottom). Thus, the mobile charge in this metal dominates the screening process, but is unable to screen out entirely the polarization-induced field (which increases further if the number of mobile electrons is reduced by hole injection; see Methods). The difference between the non-CS and CS conduction densities (Fig. 4, top and center panel) clearly shows, first, the local self-screening within each 5-layer block already observed in the bulk case; and second, a net charge imbalance – with negative and positive carriers accumulating, respectively, at the right (BZO-227/Bi-5517) and left (Bi-5517/BZO-227) interfaces – that acts against the polarization-generated field. While the overall self-screening response is incomplete, it is still amply sufficient (see Methods) to stabilize the *mono-domain* polar state even under such unfavorable electrical boundary conditions. (Due to the $\nabla D = \rho_{free} = 0$ condition across the interface, a mono-domain superlattice of thin insulating ferroelectrics is generally unstable.) Indeed, explicitly relaxing the non-CS superlattice, we find that the Bi-5517 layer is almost identical to the polar bulk phase, which confirms the stability of the mono-domain configuration.

As just shown, even a unit cell of single-domain Bi-5517 *is* polarized *and* sustains a polarization-generated field. Bi-5517 thus contradicts the natural assumption that, in the nanometric-film limit[6], polarization can never survive its own depolarization field. Indeed, Bi-5517 stands apart from any other known ferroelectric material due to the coexistence of a localized and strong polar instability (driven by the formation of Bi-O bonds) and a self-screening mechanism that cannot prevent the chemically-driven polar distortion *but* largely cancels the corresponding depolarizing field. In this context, Bi-5517 is akin to so-called hyper-ferroelectrics[18], whereby soft LO phonons are associated to a large high-frequency dielectric constant and small Born dynamical charges (a feature generally barred by the large gaps and Born charges characteristic of prototypical ferroelectric perovskites). In this sense, Bi-5517 behaves as a limiting case of hyper-ferroelectric, and might thus be considered an instance of self-screened hyper-ferroelectric metal.

In conclusion, we have designed a novel Bi-based layered-perovskite titanate that presents native metallicity – in the form of a conductive low-dimensional electron gas – and, simultaneously, complies with the requirements of a regular switchable ferroelectric. Besides its high conceptual significance, and the great fundamental interest of further characterizing the behavior of Bi-5517 and related materials, our finding opens interesting perspectives for innovative applications. Intriguing possibilities range from the fields of photovoltaics (as a metal, Bi-5517 can be expected to be a very good absorber that, simultaneously, features a built-in driving force to separate electrons and holes) to electronics (Bi-5517 may be expected to behave as a heavily n-doped semiconductor strongly responsive to applied fields) or spintronics (there are obvious strategies to construct a spin-polarized, multiferroic version of our ferroelectric metal), promising to generate great excitement around our discovery.

**Supplementary Information** is linked to the online version of the paper at www.nature.com/nature.

**Author contributions**

A.F. and P.D led the work to characterize the electric, electronic and transport properties of $Bi_5Ti_5O_{17}$. V.F. and F.R. led the work to characterize the behavior of finite layers and slabs, and the effects of doping. J.Í. led the work to identify $Bi_5Ti_5O_{17}$ as a candidate ferroelectric metal.

**Author information**

Reprints and permissions information is available at www.nature.com/reprints. The authors declare no competing financial interests. Correspondence and requests for materials should be addressed to: A.F. (alessio.filippetti@dsf.unica.it), V.F. (vincenzo.fiorentini@dsf.unica.it) and J.Í. (jorge.iniguez@list.lu).

**Acknowledgments:** Work supported in part by MIUR-PRIN 2010 project Oxide (AF, VF, PD, FR), Fondazione Banco di Sardegna (AF, VF, PD), FNR Luxembourg Grant No. FNR/P12/4853155/Kreisel (JI), MINECO-Spain Grant No. MAT2013-40581-P (JI), CINECA-ISCRA grants (AF, VF, FR, PD), CAR of UniCagliari (VF). JI ran calculations at the CESGA supercomputing center. AF, VF, PD, and FR ran calculations at CINECA.

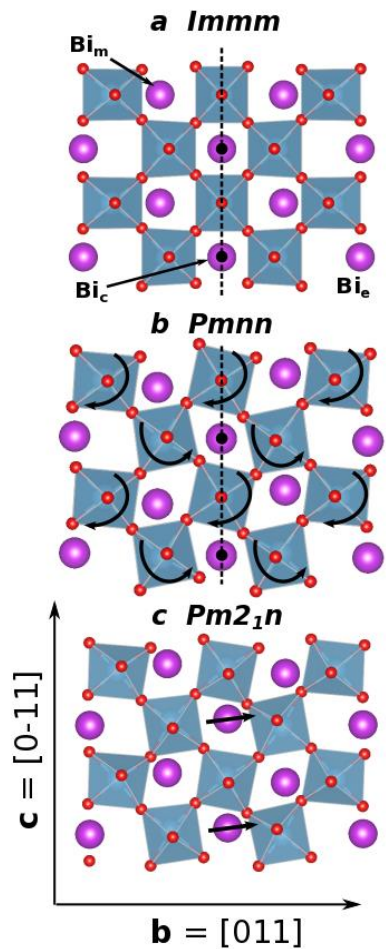

**Figure 1 | Structural detail of $Bi_5Ti_5O_{17}$ phases.** Sketch of the Bi-5517 phases discussed in the text, showing relevant structural features. **a** *Immm*, **b** *Pmnn*, **c** *Pm2$_1$n*. Only one **b**-oriented slab is shown (there are 2 such slabs in the simulation cell). Bi types indicated in **a**. **b**-oriented mirror plane, and inversion centers at the Bi$_c$ positions, indicated in **a** and **b**. In the *Pmnn* case, this mirror plane has an associated ½(**a**+**b**+**c**) glide translation; hence, the symmetry is not obvious from the figure. Bi$_c$ off-centering displacements indicated in **c**. The Bi$_c$ displacement along **c** is compensated by with a symmetric one occurring in the second slab in the cell (not shown); the Bi$_c$ displacement along **b** adds up with the symmetric one in the second slab, and thus a **b**-oriented polarization appears.

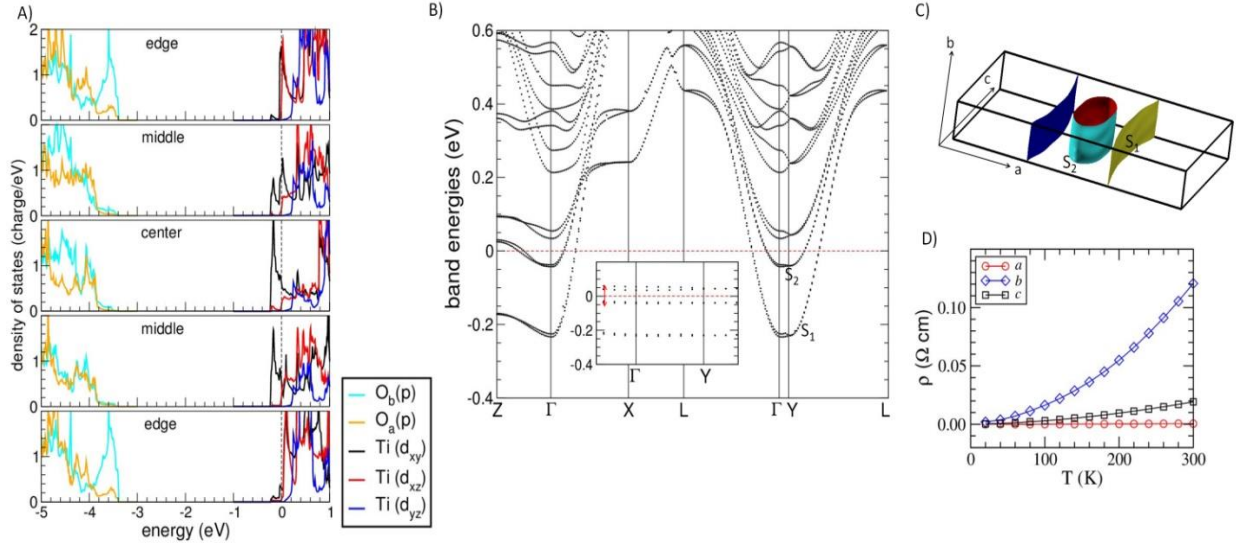

**Figure 2 | Main features of the electronic structure of $Bi_5Ti_5O_{17}$. a** Orbital- and atom-resolved DOS of $Pm2_1n$ Bi-5517. The 5 panels correspond to the 5 Ti layers in one slab. $O_a$ is in the TiO (011) planes, $O_b$ in the intermediate planes. For the $t_{2g}$ orbitals, $x$, $y$ and $z$ correspond to [100], [010] and [001], respectively. **b** Conduction band structure. Inset: zoom along $\Gamma$-Y. S1 and S2 label the occupied conduction bands, and the red-arrowed line the gap ($\Delta E = 0.1$ eV) along $\Gamma$-Y for $k_x = 0$. **c** Fermi surfaces. **d** Resistivity along the supercell axes.

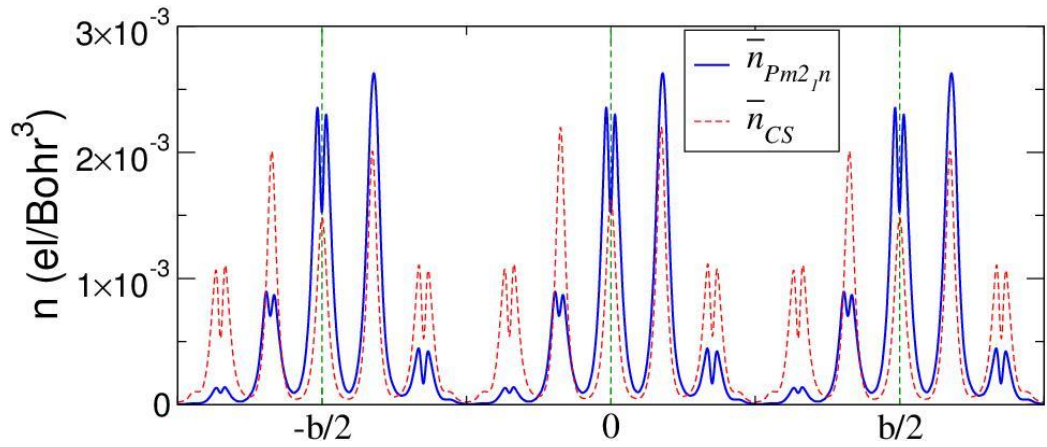

**Figure 3 | Conduction charge density of $Bi_5Ti_5O_{17}$.** The density is the result of a planar average along the stacking direction (b is the lattice constant). $\bar{n}_{CS}$ is calculated for a fictitious CS system with the cell of $Pm2_1n$ and the relative atomic positions of *Immm*. Density interpolated on a very dense grid to reduce real-space integration errors. The polarization in the $Pm2_1n$ phase points to the right.

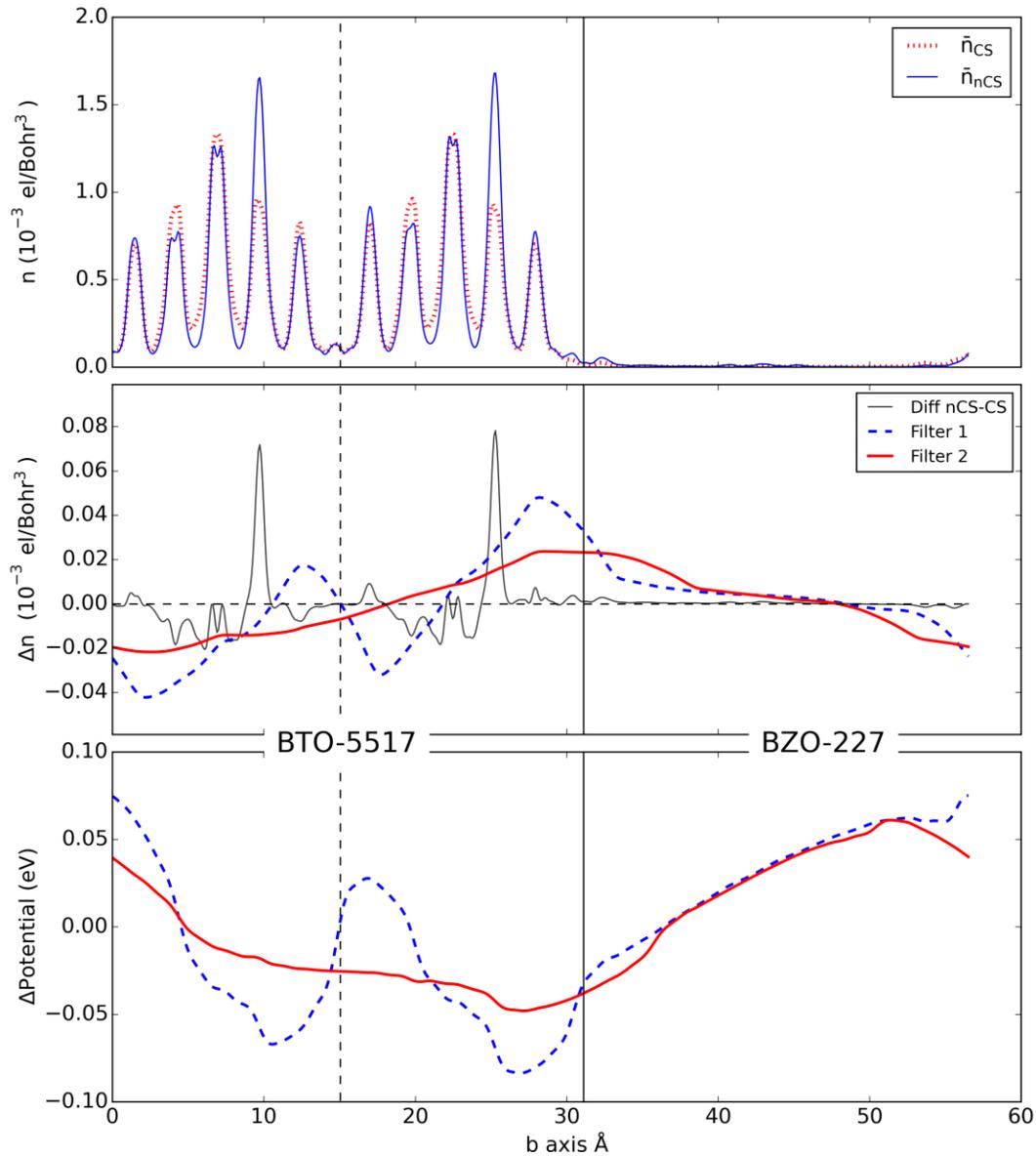

**Figure 4 | Potential and conduction charge density of $Bi_5Ti_5O_{17}/Bi_2Zr_2O_7$ superlattices.** Bi-5517 is on the left, BZO-227 on the right, separated by vertical solid lines; dashed vertical lines mark the two Bi-5517 blocks. Horizontal line in center panel is the zero density baseline. Top panel shows the planar-averaged conduction density of SL with non-CS (blue solid line) and CS (red dashed) Bi-5517. Center panel shows density differences of SL between the non-CS and CS cases in three variants: planar-averaged (gray, thin, solid) reduced by a factor 10; filtered with filter $[2d,3d]$ (labeled 2; red, thick, solid); filtered with filter $[2d,d]$ (labeled 1, blue, thick, dashed). See Methods for details. Bottom panel: filtered difference of the non-CS and CS potentials (same filters and line convention as center panel). Note that the substantial (2 eV, see Methods) interface band offset of Bi-5517 to BZO-227 cancels out in the potential difference (bottom panel), but leads to efficient confinement (top panel) of the conduction charge inside Bi-5517. The polarization of the non-CS phase points to the right.

## Methods

### Simulation details

Our calculations are performed at the first-principles level within the local density approximation[19,20] to density-functional theory and the projector-augmented wave scheme[21] to treat the interaction between ionic cores and valence electrons, as implemented in the first-principles package VASP[22-25]. The following electrons are explicitly considered in the simulations: Ti 3s, 3p, 3d and 4s; La's 5s, 5p, 5d and 6s; Bi 5d, 6s and 6p; O 2s and 2p. The electronic wave functions are represented in plane-wave basis truncated at 500 eV. For all self-consistency and force calculations, Brillouin zone integrals are computed on the k-point 6×1×5 grid, reflecting the elongated shape of the cells (a≈3.9Å, b≈31.0 to 55 Å, and c≈5.4 Å; details are given in Extended Data Table 1.)

For the analysis of the electronic structure, we also use the variational pseudo-self-interaction-corrected (VP- SIC) density-functional approach[26,27], which uses ultra-soft pseudopotentials[28] with plane-wave cutoff 476 eV. Density of states calculations use 12×4×8 grids for Brillouin zone integration.

For transport, we use two approaches: i) the Bloch- Boltzmann transport theory[29] as implemented in the BoltzTraP code[30], interpolating the band structure over a 30×14×22 ab-initio calculated k-point values; ii) an effective-mass band model which allows an easy inclusion of localized states below the mobility edge. In both cases, the relaxation time needed by Bloch-Boltzmann is calculated by analytic modeling, including the most important scattering contributions (i.e. electron-phonon and impurity scattering). The model was previously applied to describe several low-dimensional systems involving titanates, with satisfactory results[31-34].

### Structural details

For our simulations of bulk $La_5Ti_5O_{17}$ and bulk $Bi_5Ti_5O_{17}$, we employ the 54-atom cell depicted in Extended Data Fig. 1(A), which contains 2 slabs with n=5 [011]-oriented perovskite-like planes *per* slab. This cell is the conventional one for the high-symmetry reference structure (*Immm* space group) depicted in Extended Data Fig. 1a, and the primitive cell for the lower-symmetry structrues (*Pmnn* in Extended Data Fig. 1b and $Pm2_1n$ in Extended Data Fig. 1c) discussed in this work. Note that, in the *Immm* phase, the two slabs are related by a (**a**+**b**+**c**)/2 translation, where ***a***, ***b***, and ***c*** are the lattice vectors indicated in Extended Data Fig. 1a. The distortion connecting the *Immm* and *Pmnn* structures essentially consist of rotations of the $O_6$ octahedra about the ***a*** axis, as can be appreciated by comparing Extended Data Figs. 1a and 1b, and is indicated in Fig. 1 of our article. The phase of the rotations is shifted from one slab to the other; as a consequence, the primitive cell doubles. Then, the distortion connecting the *Pmnn* and $Pm2_1n$ structures has a polar character, **b** being the polar axis. The corresponding polarization is largely dominated by the off-centering of the Bi cations at the central layer of the slabs, as discussed in the main paper. Additionally, by comparing Extended Data Figs. 1b and 1c, in-plane displacements of the cations, along the ***c*** direction, can be easily appreciated. Symmetrywise, such displacements constitute a ***c***-polarized distortion within each slab; however, as can be seen in Extended Data Fig. 1c, these distortions change sign when we move from one slab to the next, the net effect being null. (Thus, along **c** we have an anti-ferroelectric pattern of sorts.) For completeness in Extended Data Table 1 we report some structural and energetic details of the various structures in the two materials.

**Berry phase polarization for metals with strongly confined conduction charges**

To compute the electronic polarization contribution from the conduction electrons we use a modified Berry phase[16,17,35], approach. In its standard version this approach shows that polarization in crystals is the integrated current flowing through the system as atoms displace from the centrosymmetric ($\lambda=0$) to the non-centrosymmetric phase ($\lambda=1$):

$$\Delta P = -\frac{2e}{8\pi^3} \int_{A_k} dk_\perp \left( \phi^{(1)}(k_\perp) - \phi^{(0)}(k_\perp) \right)$$

where $k_\perp$ spans the Brillouin zone section, of area $A_k$, orthogonal to the olarization direction, and $\phi$ is the Berry phase of the Bloch wavefunctions:

$$\phi^{(\lambda)}(k_\perp) = \lim_{N\to\infty} \Im \left\{ ln \prod_{j=0}^{N-1} \det M_{j,\nu}^{(\lambda)} \right\}$$

with

$$M_{j,\nu}^{(\lambda)} = \left\langle u_{nk_j}^{(\lambda)} \middle| u_{nk_{j+1}}^{(\lambda)} \right\rangle$$

where the $u_{nk_j}$'s are periodic parts of the Bloch wavefunctions, $\nu$ the number of bands, $n=[1,\nu]$ and $m=[1,\nu]$ band indexes, and $k_j$ runs over a string of $N$ discrete points from $\Gamma$ to $G_\parallel$, that is the shortest $G$-vector in the direction parallel to the polarization.

In general, the above expressions cannot be applied to a metal, since the Berry phase is well defined (i.e., gauge-dependent by unitary rotation) as long as the bands which contribute to the matrix in the above equation are an isolated subgroup (typically the whole valence band manifold of an insulator). Clearly for a metal this condition does not hold, since $\nu$ depends on $k$, i.e., the number of occupied bands change with $k$. However, if the system is has flat bands along a specific direction (this applies to Bi-5517 since electrons are quite localized within each 5-unit block along the ***b*** axis), then the number of occupied bands only changes with $k_\perp$, but not along the string, i.e., *no band crosses the Fermi level along the k-space string parallel to the insulating direction*. It follows that $\phi^{(\lambda)}(k_\perp)$ is well defined for any $k_\perp$, while $\nu$ can change with $k_\perp$ with the constraint

$$Q = \frac{2e}{A_k} \int_{A_k} dk_\perp \, \nu(k_\perp)$$

on the total electron charge $Q$. In practice $\phi^{(\lambda)}(k_\perp)$ may fluctuate widely with $k_\perp$ due to the change in the number of bands contributing to the string, and in turn the 2D average in the above equation may be slowly convergent. The considerable computing effort required to calculate the Berry phase in large-size systems (such as Bi-5517) suggests adopting a strategy to minimize the contribution of the variable-band-number part. In the specific case of Bi-5517, the latter contribution is limited to a few conduction bands, well separated by a large band gap from the valence bands. We therefore first calculate separately the polarization due to valence bands, which typically converges rapidly over a limited set of $k_\perp$ points. For the conduction bands, the polarization is then calculated as the 2D average of the renormalized phase

$$\bar{\phi}^{(\lambda)}(k_\perp) = \frac{Q_{\text{cond}}}{2e\nu_{\text{cond}}(k_\perp)}\phi^{(\lambda)}(k_\perp)$$

where $Q_{\text{cond}}$ is the conduction charge, and $\nu_{\text{cond}}$ the number of occupied conduction bands at $k_\perp$. With this choice, each $k$-string contribution is renormalized to the same number of electrons, and fluctuates much less with $k_\perp$, reducing the effort needed to converge the calculation.

**Depolarizing field: geometry and confinement**

The main issue in the assessment of the depolarizing field in Bi-5517 is the termination of the finite system. We checked that a slab of a normal insulating ferroelectric (PbTiO$_3$ in the (001) direction) in vacuum has the same depolarizing field with symmetric or asymmetric surface terminations. This need not apply to Bi-5517, given the presence of free charge. The symmetric BiO surface-terminated slab, obtained adding vacuum above and below the primitive cell, has essentialy zero residual field, i.e no macroscopic dipole; this is due to the conduction charge being almost entirely located at surface states, and therefore screening the polarization charge very effectively. The asymmetric termination with a TiO$_2$ layer on one side and a BiO on the other leads to a large field: however, this results again from conduction charge bound into surface states on the Ti-terminated surface, rather than from screened polarization. This is also confirmed by calculations in another asymmetric surface termination, whereby we find a large field *opposite* to that expected from polarization.

This motivates us to extract the properties of a finite Bi-5517 system using a Bi-5517/insulator superlattice. We choose Bi$_2$Zr$_2$O$_7$ (BZO-227) as our cladding insulator. This material is non-polar in the configuration we impose on it, although it turns out to be ferroelectric with polarization along the *c* axis when relaxed as stand-alone bulk. BZO is also BTO-stoichiometric on the A site, thus virtually eliminating the possibility of interface-state related fields.

BZO-227 provides a good confinement of the conduction electrons within Bi-5517, which is desirable in the present case. Examining the locally-projected density of states (DOS), we find that at the BTO/BZO interface there is essentially zero valence band offset, but a sizable conduction band offset. As shown in Extended Data Fig. 2, the centroid of the *d* DOS of the Zr adjacent to the interface is about 2 eV higher than the *d* DOS of the Ti just opposite to it across the interface. (We consider the *d* DOS since the conduction band is mostly of *d* character of the octahedrally coordinated cation.) There is a tail of Zr DOS overlapping Ti's DOS, due to a single orbital presumably involved in bonding across the interface. Thus, it seems fair to extract the conduction offset by a linear extrapolation of the two DOS to zero following the slope on the low energy side of the main peaks. This gives an offset of about 2.0 to 2.2 eV. In Extended Data Fig. 4 of the main paper this offset is not directly visible as it cancels out in the potential difference (bottom panel, Fig. 4 of main paper), but is indirectly visible in the confinement of the conduction charge inside Bi-5517 (top panel, same Figure).

**Filters**

To analyze the potential and charge density in the SLs, we apply well-known averaging processes, described e.g. in Refs. 7 and 36. The filter average features prominently in Fig. 4 of the main paper, and is defined as

$$\bar{\bar{n}}(z) = \frac{1}{L}\int_{z-L/2}^{z+L/2} dz' \int_A n(x,y,z')\,dxdy \equiv \frac{1}{L}\int_{z-L/2}^{z+L/2} dz'\,\bar{n}(z')$$

with $n$ a function (e.g. the charge density). It is a one-dimensional square-wave filter of the planar average (implicitly defined by the second equality, and calculated over a sectional area $A$) over a window of fixed width $L$, which is its defining parameter. If $L$ is the microscopic periodicity length (assuming such periodicity exists), all microscopic oscillations in the planar average are eliminated. This basically amounts to filtering away all the Fourier components corresponding to microscopic oscillations.

However, the typical potential or charge density in our system has a rather wide spectrum in wavevector space, and filtering all components would entail a complete loss of information. So, in practice we choose filters such that microscopic oscillations are reduced significantly after a couple of passes at most. We find that applying the filter twice, with $L=d$ and $2d$, or with $L=3d$ and $2d$, gives the best results in terms of oscillation removal, $d$ being the average interplanar distance.

**Depolarizing field vs density of mobile charge**

Any reduction of the conduction charge (2 electrons per cell in Bi-5517) should cause an increase of the depolarizing field. This reduction may be produced by doping or field-effect injection; for example, Ca substitution of Bi at the 10 % level (i.e. two Bi-5517 units become $CaBi_9Ti_{10}O_{34}$) will reduce the conduction charge in the unit cell to 1 electron. Since doping is difficult to simulate and may lead to unintended consequences (such as modifying the polarization, etc.), we study the effect of conduction charge removal for the same superlattice as before, simply subtracting by hand a certain amount of charge $\Delta Q$. (The ions are kept clamped for simplicity; as mentioned in the main paper, the polar distortion is unaffected by relaxation.) The charge is effectively removed from within Bi-5517, and neutrality is maintained by a uniform compensating background that spreads over the whole cell. The spurious potential thus induced is eliminated automatically by taking the difference of the potentials of non-CS and CS superlattices. We can thus compare the filtered averages of the potential for various values of $\Delta Q$ all the way up to 2 electrons, i.e. to zero conduction charge remaining, and extract the field in the superlattice as a function of removed charge.

In Extended Data Fig. 3 we report the potential difference between non-CS and CS superlattices for various values of $\Delta Q$, for both the filters mentioned in the previous Section, in the two top panels. The [$d,2d$] filter, in particular, highlights the local screening within each block of Bi-5517. In the same Figure, bottom panel, we show the fields extracted from the total potential drop across the gray-shaded regions for both filters. Clearly, the values are quite similar in both cases. In the limiting case of 2 electrons removed, the field is limited by the gap to about 0.5 GV/m. The valence electron screening brings it down further to about 0.1 GV/m, for an effective valence dielectric constant of 5.

**Ferroelectricity in finite field**

It is known[37] that the energy associated with the depolarizing field may destroy ferroelectricity in a finite system. We checked that is the case for a PbTiO$_3$ slab containing four Ti units. The unscreened bulk polarization charge (0.8 C/m$^2$) would generate a field of 90 GV/m and a field energy of 100 eV in the simulation cell. With ions clamped in the ferroelectric configuration, the purely electronic response reduces the field to 5 GV/m, i.e. a 0.3 eV field energy. Since the ferroelectric well depth of PbTiO$_3$ is about 0.08 eV per Ti, this field energy is sufficient to remove ferroelectricity. Indeed relaxing the ions we recover the paraelectric geometry.

This is not an issue for Bi-5517. The bare $\Delta P$ field is 37 GV/m (field energy 51 eV in the SL cell), but the screened field is 0.02 GV/m, and hence the field energy becomes negligible and the system is comfortably on the ferroelectric-stable side.

**Resistivity upturn at low *T***

Figure 2D of the main text shows that the resistivity is anisotropic and metal-like. Resistivity experiments[8] for La-5517 (nearly isostructural to our newly designed Bi-5517, and the closest existing material in many respects[37]) suggest that the resistivity components are insulator-like over significant temperature intervals. The resistivity is activated, with several different small activation energies at play, indicating that the upturn is probably due to defects of some kind. This situation can be described theoretically considering an effective-mass two-band model including a light-mass (0.7 $m_e$) $t_{2g}$ band edge and a single localized state lying 40 meV lower, contributing to current by thermal activated hopping. The calculated resistivities along the crystal axes, shown in Extended Data Fig. 4, are anisotropic and have an insulating upturn quite similarly to experimental ones. This corroborates the attribution of the insulator-like resistivity upturn in this low-density, flat Fermi-surface metal to low-activation-energy defects.

**Additional references for the Methods section:**

19. **Ceperley, D. & Alder, B. J.** Ground State of the Electron Gas by a Stochastic Method. Phys. Rev. Lett. **45**, 566 (1980).

20. **Perdew, J. P. & Zunger, A.** Self-interaction correction to density-functional approximations

portance of being multiple-band conductors. Phys. Rev. B **86**, 195301 (2012).

32. **Delugas, P., Filippetti, A., Verstraete, M. J., Pallecchi, I., Marré, D. & Fiorentini, V.** Doping-induced dimensional crossover and thermopower burst in Nb-doped $SrTiO_3$ superlattices. Phys. Rev. B **88**, 045310 (2013).

33. **Delugas, P., Filippetti, A., Gadaleta, A., Pallecchi, I., Marré, D. & Fiorentini, V.** Large band offset as driving force of two-dimensional electron confinement: The case of $SrTiO_3/SrZrO_3$ interface. Phys. Rev. B **88**, 115304 (2013).

34. **Pallecchi, I., Telesio, F., Li, D., Fête, A., Gariglio, S., Triscone, J.-M., Filippetti, A., Delugas, P., Fiorentini, V. & and Marré, D.** Giant oscillating thermopower at oxide interfaces. Nature Commun. 6:6678; doi: 10.1038/ncomms7678 (2015).

35. **Vanderbilt, D. & and King-Smith, R. D.** Electric polarization as a bulk quantity and its relation to surface charge. Phys. Rev. B **48** 4442 (1993).

36. **Peressi M., Binggeli N. and Baldereschi A.** Band engineering at interfaces: theory and numerical experiments. J. Phys. D: Appl. Phys. **31**, 1273 (1998).

37. **Meyer, B. & and Vanderbilt, D.** Ab initio study of $BaTiO_3$ and $PbTiO_3$ surfaces in external electric fields. Phys. Rev. B **63**, 205426 (2001).

38. **Nanamatsu, S., Kimura, M., Doi, K., Matsushita, S. & Yamada, N.** A new ferroelectric: $La_2Ti_2O_7$. Ferroelectrics **8**, 511 (1974).

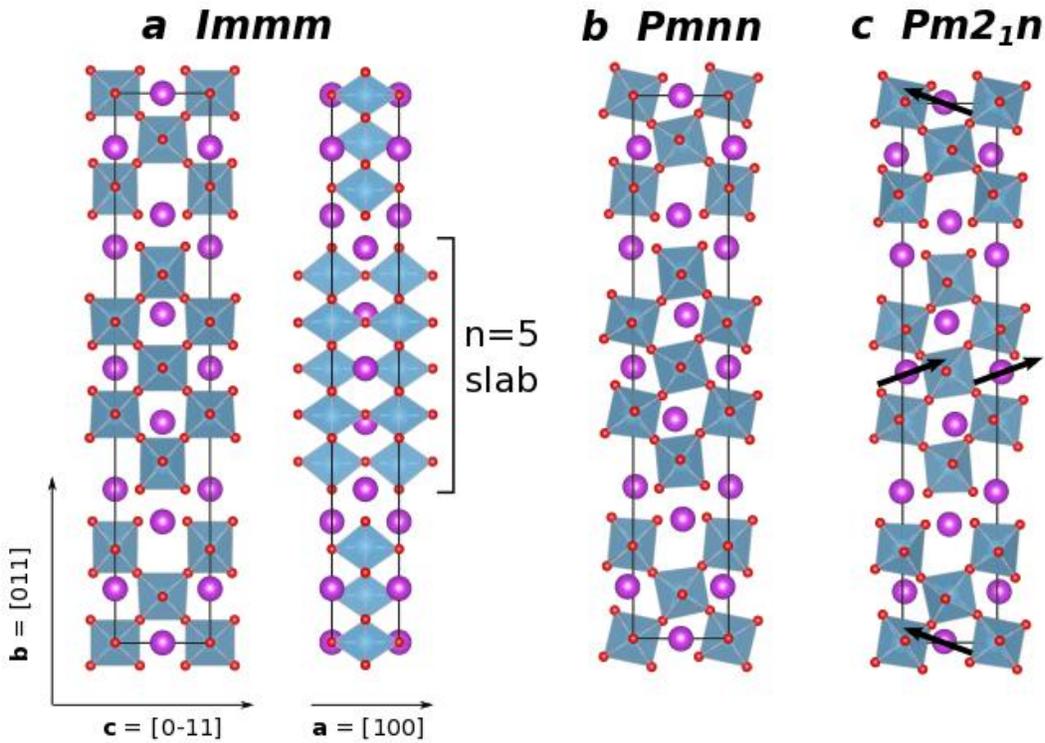

**Extended Data Figure 1 | Structural detail of the $Bi_5Ti_5O_{17}$ simulations.** We show the *Immm* (**a**), *Pmnn* (**b**), and *Pm2$_1$n* (**c**) structures of layered n=5 perovskite, as obtained from first-principles relaxations. The directions of the supercell vectors are indicated in **a**; the Cartesian setting is chosen to coincide with the pseudo-cubic axes of the perovskite lattice. In **c** we indicate the off-centering displacement of the Bi cations in the central layers.

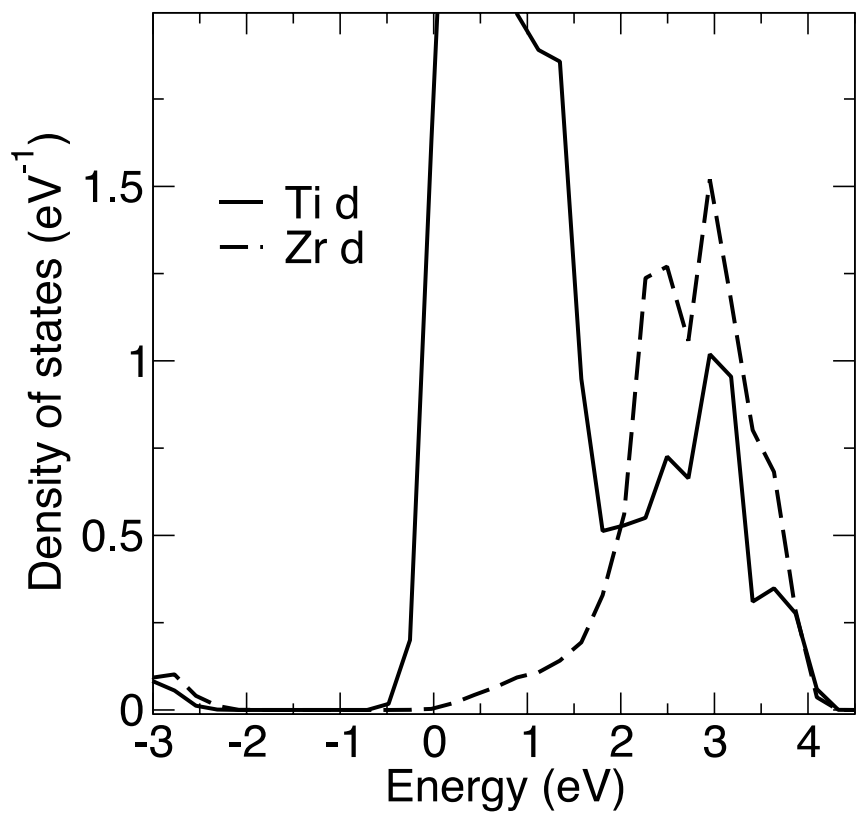

**Extended Data Figure 2 | Local density of states of the $Bi_5Ti_5O_{17}/Bi_2Zr_2O_5$ superlattice.** We show the projections on the Ti and Zr atoms adjacent to the interface on the right side of Fig. 4 of the main text (the low potential side).

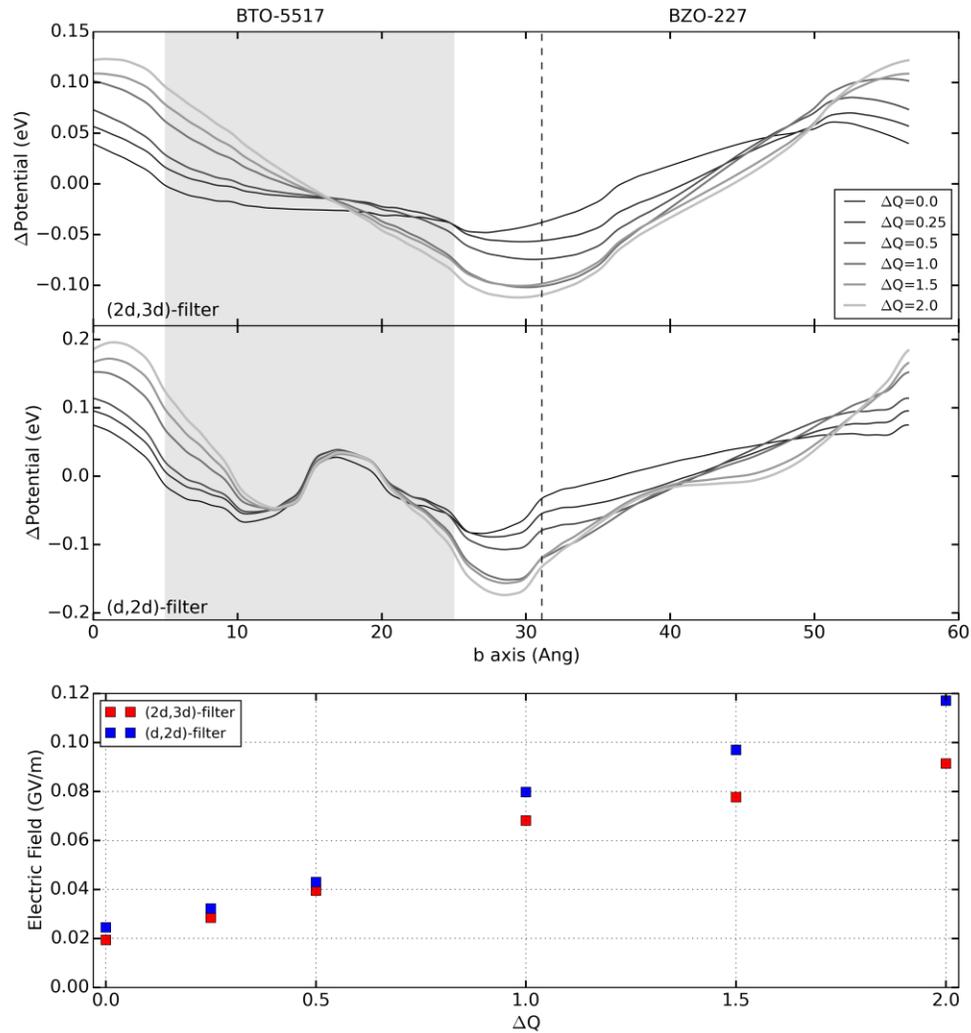

**Extended Data Figure 3 | Superlattice results as a function of hole doping.** Top and center panels: filter averages of the potential difference between the non-CS and CS states of the Bi-5517/BZO-227 superlattice for removed charges from 0 to 2 (the shade of gray is lighter and the thickness increases along the series). Clearly the slope increases with the removed charge. Bottom panel: field value as a function of the removed charge for the two different filters.

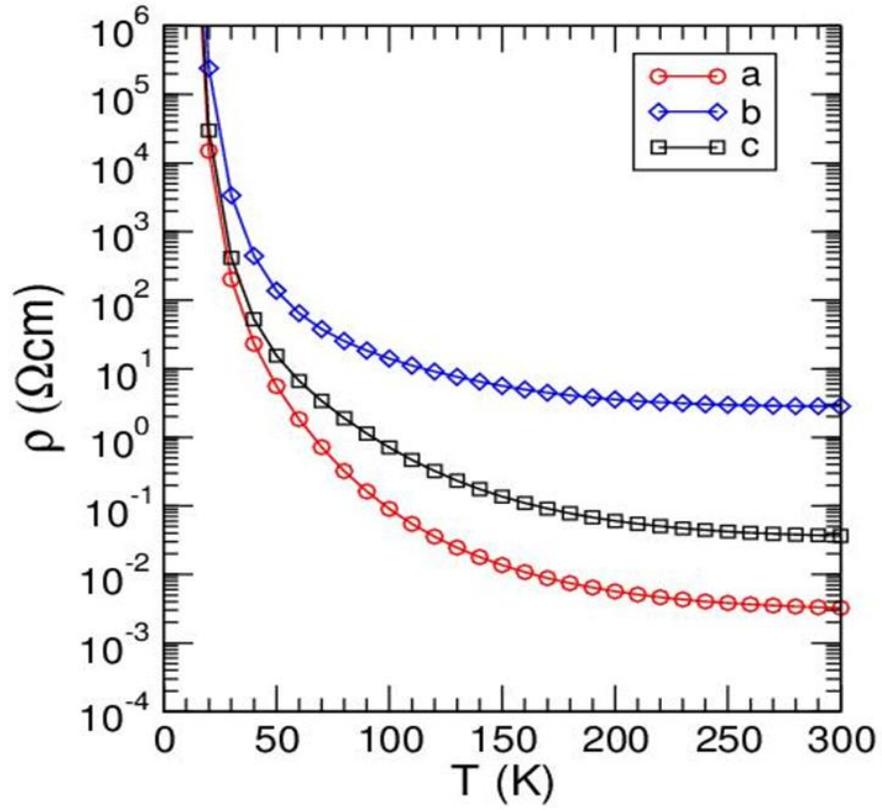

**Extended Data Figure 4 | Resistivity in presence of defects.** We calculate it by an effective-mass model (see text) along the crystallographic axes of Bi-5517.

|       | La-5517 |       |        |       | Bi-5517 |       |        |       |
|-------|---------|-------|--------|-------|---------|-------|--------|-------|
| Phase | E       | $a$   | $b$    | $c$   | E       | $a$   | $b$    | $c$   |
| $Immm$  | 0.74  | 3.885 | 31.276 | 5.439 | 1.55    | 3.863 | 31.461 | 5.435 |
| $Pmnn$  | 0     | 3.912 | 30.805 | 5.422 | 0.21    | 3.902 | 30.975 | 5.418 |
| $Pm2_1n$ | –    | –     | –      | –     | 0       | 3.890 | 31.100 | 5.461 |
| Expt  |         | 3.929 | 31.466 | 5.532 | –       | –     | –      | –     |

**Extended Data Table 1 | Details of investigated phases.** Energy (eV/cell) and lattice constants (Å) of various phases of La-5517 and Bi-5517. $a$, b, and $c$ are the lengths of the crystal vectors, which are parallel to the [100], [011], and [0$\bar{1}$1] directions, respectively. The ground states are *Pm2$_1$n* for Bi-5517 and *Pmnn* for La-5517. Other energies are referred to these ground states. Experimental lattice parameters[8] for La-5517 are also reported for comparison.